\documentclass[%
 reprint,
 amsmath,amssymb,
 aps,
]{revtex4-2}

\usepackage{graphicx}
\usepackage{dcolumn}
\usepackage{bm}

\usepackage{multirow}
\usepackage{subfigure}

\begin{document}

\preprint{APS/123-QED}

\title{Hypergeometric viable models in $f(R)$ gravity}

\author{Roger A. Hurtado}
 \email{rahurtadom@unal.edu.co}
\author{Robel Arenas}%
 \email{jrarenass@unal.edu.co}
\affiliation{%
Observatorio Astronómico Nacional, Universidad Nacional de Colombia, Bogotá, Colombia
}

\date{\today}

\begin{abstract}
A cosmologically viable hypergeometric model in the modified gravity theory $f(R)$ is found from the need for asintoticity towards $\Lambda$CDM, the existence of an inflection point in the $f(R)$ curve, and the conditions of viability given by the phase space curves $(m, r)$, where $m$ and $r$ are characteristic functions of the model. To analyze the constraints associated with the viability requirements, the models were expressed in terms of a dimensionless variable, i.e. $R\to x$ and $f(R)\to y(x)=x+h(x)+\lambda$, where $h(x)$ represents the deviation of the model from General Relativity. Using the geometric properties imposed by the inflection point, differential equations were constructed to relate $h'(x)$ and $h''(x)$, and the solutions found were Starobinsky (2007) and Hu-Sawicki type models, nonetheless, it was found that these differential equations are particular cases of a hypergeometric differential equation, so that these models can be obtained from a general hypergeometric model. The parameter domains of this model were analyzed to make the model viable.
\end{abstract}

\maketitle


\section{Introduction}
General Relativity (GR) is the most widely accepted theory of gravity, it predicts the expansion of the Universe and the associated redshifts of the galaxies as a dynamic consequence of its evolution from the so-called Big Bang; as well as other remarkable phenomena such as the gravitational lensing effect \cite{gravitationalenses,Planck20132}, black holes and gravitational waves \cite{einstein}, recently detected \cite{Abbott2016}. However, the conclusions of the observational data of Supernovae type Ia (SN Ia) \cite{Riess_1998,Perlmutter_1999}, showed that the Universe experiences an accelerated expansion phase, this fact has no clear interpretation in the framework of GR and it was necessary to introduce a type of unknown negative pressure force, called Dark Energy (DE), whose action would dominate gravitational attraction on large scales \cite{Perlmutter_1999,Carroll_2001,Riess:2004nr}. This model is known as Lambda Cold Dark Matter ($\Lambda$CDM), which also takes into account a new and strange type of gravitating matter, but not interacting with radiation, called Dark Matter (DM) \cite{zwicky2, Navarro_1996}, whose effect is to correct the discrepancy between the theory and the observed flat rotation curves of spiral galaxies \cite{ostriker, rotationcurves}.
\\
$\Lambda$CDM fits very well within a wide spectrum of cosmological observations, \cite{Planck20131,Planck20132,Planck20133,Planck2016,Jarosik_2011}, however, the nature of DM and DE is unknown, even though according to observations made by the ESA's Planck satellite in 2013 \cite{planck20130}, within the theoretical scope of $\Lambda$CDM, their content is 27\% and 68\% respectively, i.e. what is known of the Universe comprises only 5\% of its energy density, which raises even more questions than answers.
\\
$\Lambda$CDM is indeed a paradigm, and in the course of the last century new ideas were added to complement it, such is the case of the theory of cosmological inflation, which accounts for the homogeneity and isotropy of the Universe at large scale from the accelerated expansion of the early Universe \cite{Planck20133}, solving, among others, the flatness problem \cite{guth} and the magnetic monopole problem \cite{LINDE1982389}; however, to date there is no generally accepted model for inflation and likewise the standard model of cosmology has some problems (see \cite{problems} for a synthesis on this subject) that make it necessary to reconsider our understanding of GR on cosmological scales. One such alternative, motivated mainly by the search for a geometrical explanation for the late-time acceleration, is the $f(R)$ theory \cite{Sotiriou:2008rp,Perez_Romero_2018}, whose dynamics is obtained from an action written in terms of a general function of the scalar curvature, $R$. There are several $f(R)$ models for solving the DM \cite{Yadav_2019, Boehmer:2007kx, Capozziello:2004km}, DE \cite{Amendola_2007a,Amendola_2007b,Capozziello:2004km} problems and even the inflationary phase, whose first model in the context of $f(R)$ theory, was proposed by Starobinsky in 1980 \cite{Starobinsky:1980te}, which is constructed by adding to Einstein-Hilbert (EH) action a quadratic term for the curvature scale, i.e. $f(R)=R+\alpha R^2$, with $\alpha$ constant, this model has been carefully studied and is in agreement with the data recently observed by the Planck satellite \cite{planck2018}.

The task of finding a viable $f(R)$ model, which reproduces inflation, radiation-dominated stage followed by the matter-dominated phase and late-time accelerated expansion, while being able to pass the tests of the Solar system, is not at all easy, however in Ref \cite{Amendola_2007b} the general conditions for a model to be cosmologically acceptable are found, and examples of viable models are Starobinsky \cite{Starobinsky_2007}, Hu-Sawicki \cite{Hu_2007}, Tsujikawa \cite{Tsujikawa_2008} and exponential models \cite{Linder_2009}. In particular the first two models have been tested using redshift of SN Ia data, their cosmological and free parameters were calculated using a Markov chain Monte Carlo simulation, and it was concluded that these models fit the data with high accuracy \cite{hough2020viability}. One characteristic of these models is that they present an inflection point, this property will be discussed in this paper, focussing on the conditions that $f(R)$ models must possess in order to be considered cosmologically valid.

This work is organized as follows. In Sec. \ref{sec:2} will review the theoretical framework of $f(R)$ theory, outlining the field equations. Sections \ref{sec:3} and \ref{sec:4} will be used to analyze the conditions of viability of $f(R)$ models together with the existence of an inflection point in the function. In sections \ref{sec:5} and \ref{sec:6} a differential equation will be constructed from the geometric properties imposed by the conditions mentioned above, the solutions will be shown and generalized as a hypergeometric model in section \ref{sec:7}. The conclusions will be shown in Sec. \ref{sec:discu}.

\section{Field equations in $f(R)$ theory}\label{sec:2}
The $f(R)$ theory is constructed from a modification of the E-H action, where the Lagrangian density is an arbitrary function of $R$, defined over a hypervolume $\Sigma$ \cite{Sotiriou:2008rp}
\begin{equation}
I=\frac{1}{2\kappa}\left(\int_{\Sigma} d^4x\sqrt{-g}f(R)+I_{GYH}\right)+I_{M},
\end{equation}
where $I_{GYH}$ is the Gibbons-York-Hawking boundary ($\partial\Sigma$) term \cite{Guarnizo:2010xr}, $I_M$ is the contribution of matter, and $\kappa=8\pi G$. In the metric formalism, by varying the action with respect to the metric $g_{\mu\nu}$, the modified field equations are obtained,
\begin{equation}\label{fieldeq}
F R_{\mu\nu}-\frac{1}{2}f g_{\mu\nu}-F_{;\mu\nu}+g_{\mu\nu}F_{;\alpha}^{;\alpha}=\kappa T_{\mu\nu},
\end{equation}
where $f=f(R)$, $F=F(R)=f'(R)$, the D'Alembertian is defined by $F_{;\alpha}^{;\alpha}$, the covariant derivatives $\nabla_\mu\nabla_\nu F=F_{;\mu\nu}$ and the definition of the energy-momentum tensor
\begin{equation}
T_{\mu\nu}=-\frac{2}{\sqrt{-g}}\frac{\delta I_M}{\delta g^{\mu\nu}}.
\end{equation}
\\
The trace of the field equations (\ref{fieldeq}), is obtained by multiplying by the metric tensor
\begin{equation}\label{trace}
FR-2f+3F_{,\alpha}^{,\alpha}=\kappa T,
\end{equation}
where $T$ is defined as $T = g^{\mu\nu}T_{\mu\nu}$. Even though Eq. (\ref{trace}) is a differential equation, as in GR, usually it is taken as an algebraic equation to relate $R$, $f(R)$ and $F(R)$. In GR, $T=0$ implies that $R=0$, but this does not hold in $f(R)$ theory, in the case where there are also coupled Maxwell fields, whose stress-energy tensor is traceless, the scalar curvature is constant $R=R_0$, however when the electromagnetic fields are non-linear, for example of the Born-Infeld type, the solutions imply $R=R(r)$, where $r$ is radius vector, see Ref. \cite{Hurtado_2020}. By means of the trace, the field equations are reduced to
\begin{equation}\label{fieldequ}
    F R_{\mu \nu }-\kappa  T_{\mu \nu }-F_{;\mu \nu }-\frac{1}{4}\left(F R-\kappa T-F_{;\alpha }^{;\alpha }\right)g_{\mu \nu }=0.
\end{equation}
This equation depends on the second covariant derivatives of the scalar function $f(R)$, which are combinations of partial derivatives of the metric. The main advantage of the $f(R)$ theories of gravity is the possibility of returning to GR quickly, simply by making $f(R)=R$.

\section{Model constraints and inflection point}\label{sec:3}
The general form of the function $f(R)$ can be expressed explicitly as the sum of the linear term $R$ which reproduces GR plus a perturbation,
\begin{equation}\label{fdefinition}
f(R)=R+\tilde{f}(R)+\lambda R_0,
\end{equation}
where $\tilde{f}(R)$ represents the deviation of the model from GR, and the $\Lambda$CDM model can be obtained as a special case with $\tilde f(R)=0$, where $\lambda=-2 \Lambda/R_0$, and $\Lambda$ is the cosmological constant. Thus, when defining the dimensionless coordinate by making $x=R/R_0$,
\begin{equation}\label{xdefinition}
    y(x)=x+h(x)+\lambda,
\end{equation}
where $y(x)=f(R_0x)/R_0$, $\tilde{f}(R_0x)=R_0h(x)$, and with the definition of
the characteristic functions \cite{Amendola_2007b}
\begin{equation}\label{m}
    m=\frac{Rf''(R)}{f'(R)}=\frac{xh''(x)}{1+h'(x)},
\end{equation}
and
\begin{equation}\label{r}
    r=-\frac{Rf'(R)}{f(R)}=-\frac{x\left[1+h'(x)\right]}{x+h(x)+\lambda}.
\end{equation}
Now, let us suppose that $h(x)$ is a continuous function, with continuous derivatives in a domain $I$, such that $x_i\in I$ is an inflection point of $h(x)$, not stationary nor of infinite slope. That is, $h''(x_i)=0$, which means that $h'(x_i)$ is a maximal point. The existence of an inflection point, together with the conditions of asintoticity towards $\Lambda$CDM  \cite{Hu_2007},
\begin{equation}\label{limith0}
    \lim_{x\to0}h(x)=-\lambda,
\end{equation}
and
\begin{equation}\label{limithinfty}
    \lim_{x\to\infty}h(x)=\frac{k}{R_0}-\lambda,
\end{equation}
where $k$ is a constant that depends on the model; restrict the form of the function $h(x)$ to only two possibilities: decreasing concave (increasing convex) at $0<x<x_i$ and decreasing convex (increasing concave) at $x_i<x$. Motivated by the results of Starobinsky model \cite{hough2020viability} whose virtue lies in the quadratic term,
which reproduces accelerated expansion of the Universe without the need to introduce Dark Matter, and for $h(x)$ to contain only quadratic terms when expanded in Maclaurin series, it must be an even function; this implies that $x=0$ is a maximal point and therefore
\begin{equation}\label{limithprima0}
    \lim_{x\to0}h'(x)=0,
\end{equation}
and at the other hand at infinity the curve is flattened according to Eq. (\ref{limithinfty}), thus
\begin{equation}\label{limithprimainfty}
    \lim_{x\to\infty}h'(x)=0,
\end{equation}
and in turn
\begin{equation}\label{limithdobleprima0}
    \lim_{x\to0}h''(x)=c,
\end{equation}
where $c$ is a constant, and
\begin{equation}\label{limithdobleprinfini}
    \lim_{x\to\infty}h''(x)=0,
\end{equation}
The two possible behaviors of the function affect the sign of its derivatives, as discussed in Table \ref{tab:1}.
\begin{table}[]
\begin{tabular}{|c|c|c|c|}
\hline
                   & \multicolumn{2}{c|}{$h(x)$} &                              \\ \hline
                   & Decreasing   & Increasing   & Domain                       \\ \hline
$h'(x)$            & $-$            & $+$            & $x>0$                    \\ \hline
$h''(x)$           & $-$            & $+$            & $0<x<x_i$                \\ \hline
$h''(x)$           & $+$            & $-$            & $x>x_i$                  \\ \hline
$h'(x)-h'(x_i)$    & $+$            & $-$            & \multirow{2}{*}{$x_i>0$} \\ \cline{1-3}
$h'''(x_i)$        & $+$            & $-$            &                          \\ \hline
\end{tabular}
\caption{Sign of the first derivatives of the function $h(x)$ according to its monotonicity in the domain given in the last column.}
\label{tab:1}
\end{table}
Only models with a characteristic function, (\ref{m}), $m\geq0$ and close to $\Lambda$CDM can be considered cosmologically viable \cite{Amendola_2007b}, thus there are two options for $h(x)$ in which this can be fulfilled, $h''(x)\geq0$ and $h'(x)>-1$, or $h''(x)\leq0$ and $h'(x)<-1$; so if $h(x)$ is a decreasing function, for $x\geq x_i$,
\begin{equation}\label{condition1}
    -1<h'(x)<0,
\end{equation}
or for $0\leq x\leq x_i$, 
\begin{equation}\label{condition2}
    h'(x)<-1.
\end{equation}
From the second option, when $h(x)$ is an increasing function, it is only possible to choose
\begin{equation}\label{condition3}
    h'(x)>0,
\end{equation}
for $0<x\leq x_i$.
\begin{figure}[ht]
\centering
\includegraphics[width=0.42\textwidth]{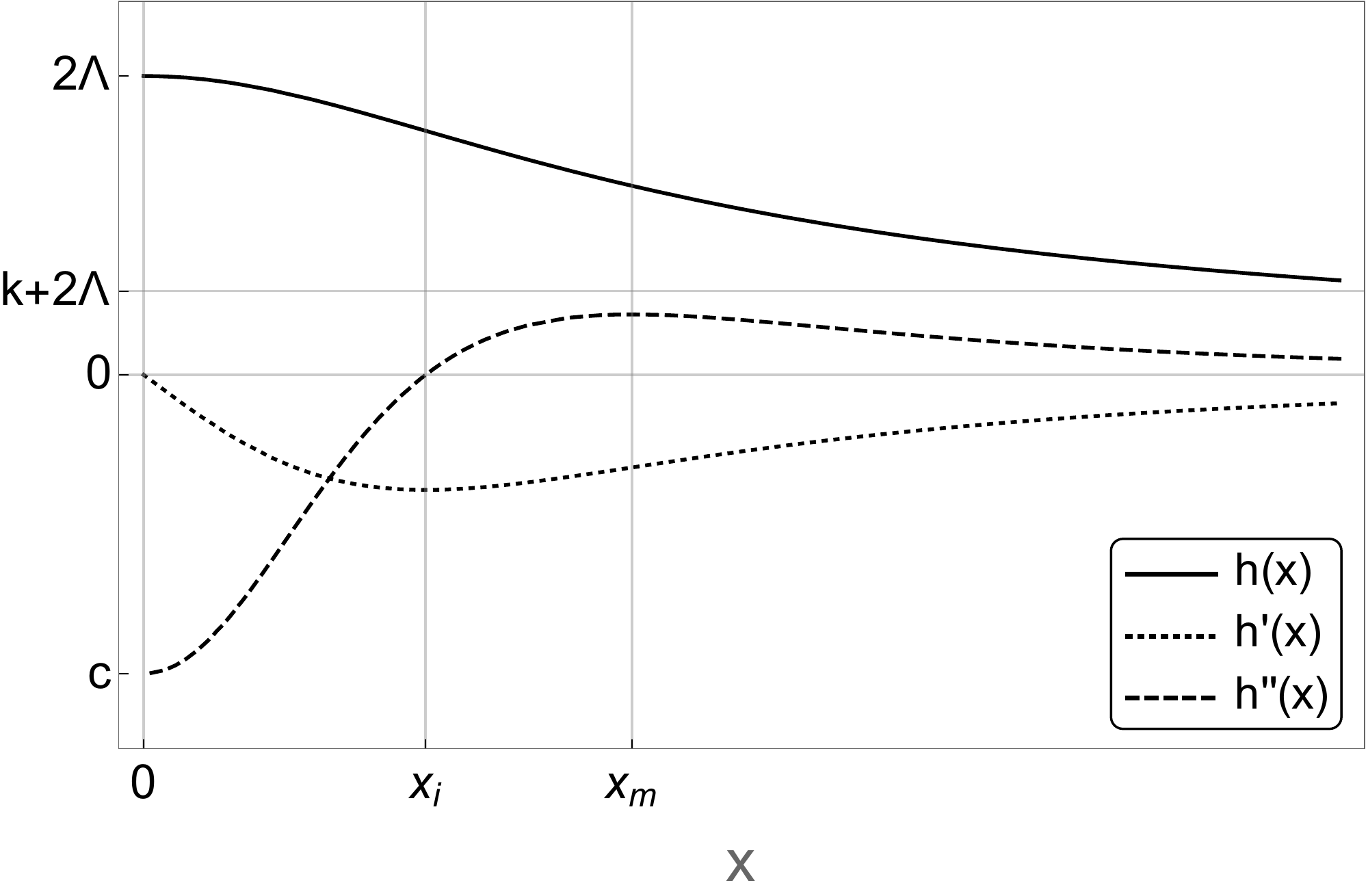}
\caption{Behaviour of the function $h$ and its derivatives as a function of $x$, with the necessary conditions for the model to be viable, equations (\ref{limith0}) to (\ref{limithdobleprinfini}), together with the existence of an inflection point, $x_i$. The image also shows the existence of the maximum of $h''(x)$, $x_m$. In this diagram $R_0=1$ was chosen.}\label{fig1}
\end{figure}
Due to condition (\ref{limithprima0}), option (\ref{condition2}) is discarded and to avoid singularities in the characteristic functions $m(x)$ and $r(x)$, option (\ref{condition3}) will also be discarded. The inflection point $x_i$ leads to the appearance of a minimum (maximum) in $h'(x)$ if $h(x)$ is a decreasing (increasing) function, and by Eq. (\ref{limithdobleprinfini}), there is an inflection point, $x_m$, on the curve of $h'(x)$ and therefore $h''(x_m)$ will be a maximal at $x>x_i$ and $h''(x)$ will be a decreasing (increasing) function for $x>x_m$. The function $h''(x)$ is integrable in all its domain, that is, by limits (\ref{limithprima0}) and (\ref{limithprimainfty}), $\int_{0}
^{\infty}h''(x)dx=0$, thus the decreasing monotonicity of $h''(x)$ from $x_m$ is the property that allows to consider the option (\ref{condition1}) as the most viable, because if $x_m<\frac{x}{2}\leq t\leq x$, then
\begin{equation}
    0\leq h''(x)\leq h''(t),
\end{equation}
so that
\begin{equation}
    0\leq \int_{x/2}^{x}h''(x)dt\leq \int_{x/2}^{x} h''(t)dt,
\end{equation}
or
\begin{equation}\label{conditionlimit}
    0\leq xh''(x)\leq 2\left[h'(x)-h'(x/2)\right].
\end{equation}
Similarly, by integrating Eq. (\ref{condition1}), it is obtained
\begin{equation}\label{eq22}
    0<x+h(x)+\lambda.
\end{equation}
Conditions (\ref{conditionlimit}) and (\ref{eq22}) are useful for calculating the limits of the $r$ function (\ref{r}), as will be seen below.
\section{Characteristic functions}\label{sec:4}
\begin{figure}[ht]
\centering
\includegraphics[width=0.42\textwidth]{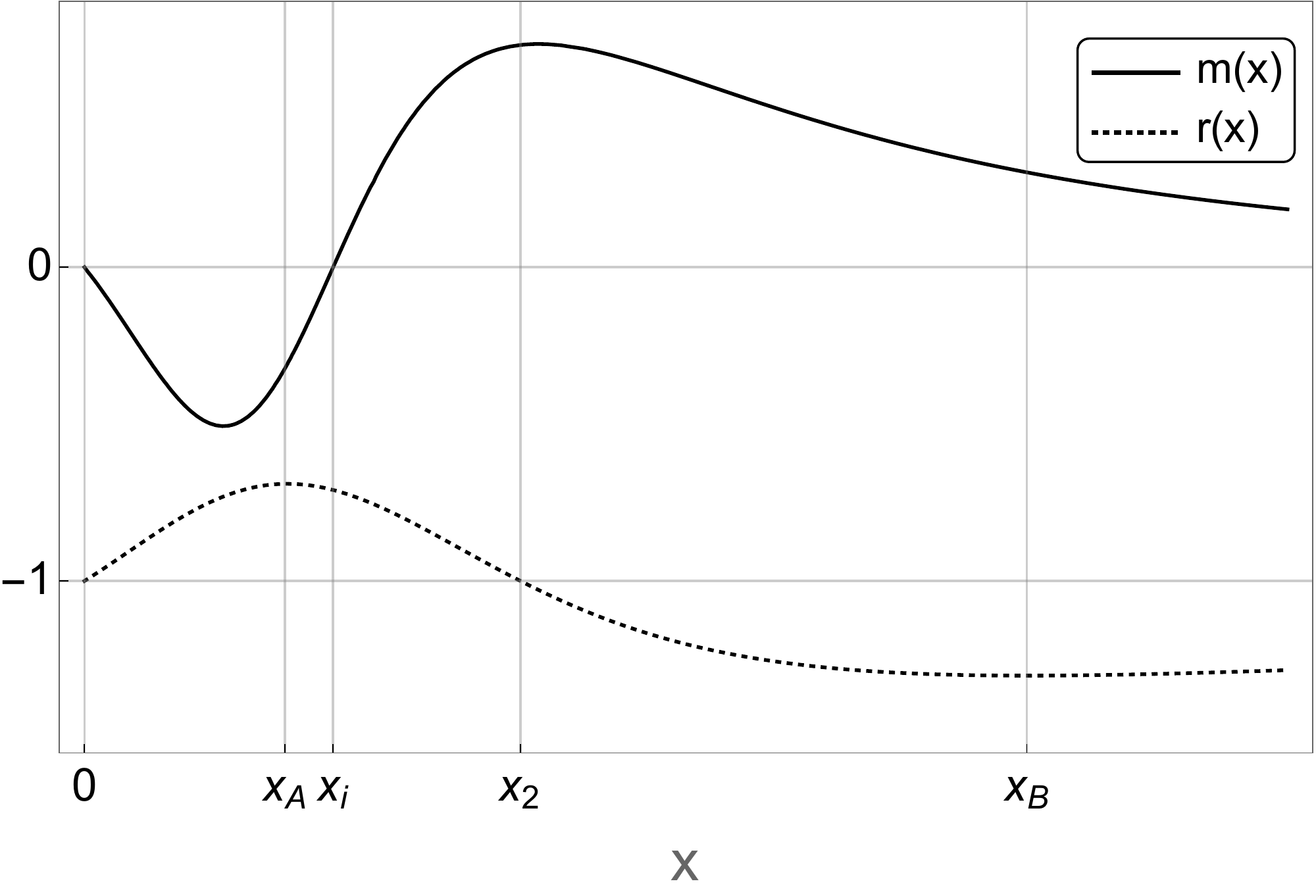}
\caption{Characteristic functions $m$ and $r$ versus $x$. The inflection point, $x_i$, the $x_2$ point, in which $r=-1$, and the maximum points of $r(x)$, $x_A$ and $x_B$, are appreciated.}\label{fig2}
\end{figure}

\begin{figure}[ht]
\centering
\includegraphics[width=0.42\textwidth]{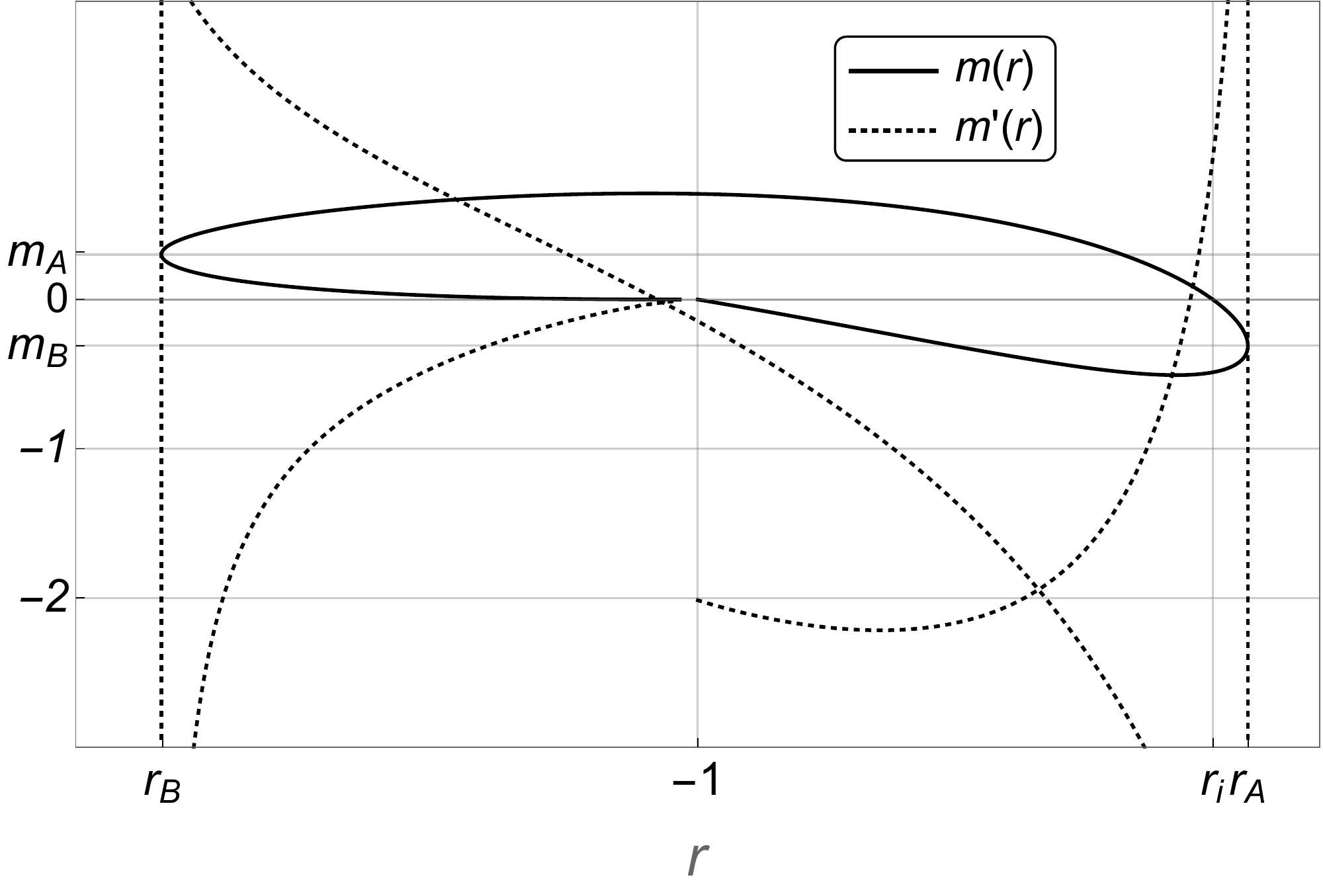}
\caption{Phase space in the plane $(m, r)$ as well as the corresponding derivative ($m'(r)$). It can be seen that in the points $r_A$ and $r_B$ the derivative is infinite. Note the behaviour of $m'(r)$ at $r(x=0)=-1$, given by Eq. (\ref{derivative2}), and at $r(x\to\infty)=-1$, the latter allowing the viability of the model.}\label{fig3}
\end{figure}

A model that reproduces a matter-dominated era with a corresponding transition to accelerated expansion must satisfy \cite{Amendola_2007b}
\begin{equation}\label{conditionsmatterdomi}
    m(r)\approx+0,\quad \text{and}\quad m'(r)>-1\quad\text{at}\quad r\approx-1,
\end{equation}
where
\begin{equation}\label{derivativem}
    \left.\frac{dm}{dr}\right|_x=\frac{x m'(x)}{r(x)[1+m(x)+r(x)]}.
\end{equation}
There are three points $x$ for which $r(x)\approx-1$, these are $x_1\to0$,
\begin{equation}\label{point1}
    \lim_{x\to0}\frac{x\left[1+h'(x)\right]}{x+h(x)+\lambda}=1, 
\end{equation}
where we have used Eq. (\ref{limith0}) and (\ref{limithprima0});
\begin{equation}\label{point2}
    x_2\to\frac{h(x_2)+\lambda}{h'(x_2)};
\end{equation}
and $x_3\to\infty$, since
\begin{equation}\label{point3}
    \lim_{x\to\infty}r(x)=\lim_{x\to\infty}\frac{1+h'(x)+x h''(x)}{1+h'(x)}=1,
\end{equation}
where L'Hospital's rule and condition (\ref{conditionlimit}) were used. Now, since $m(x_1)=-0$, and
\begin{equation}\label{derivative2}
    m'(r)|_{x_1}\overset{x\to0}{=}-2,
\end{equation}
(see Appendix A), we discard $x_1$, and noting that $m(x)$ is directly proportional to $h''(x)$, $x_i$ is also a root of $m(x)$, i.e $x_2\to x_i$, but to be a valid point, $x_2$ must tend to $x_i$ on the right, $x_{r+}$, satisfying $h(x_{r+})=x_{r+}h'(x_{r+})-\lambda$, however the last term of Eq. (\ref{derivativem}) diverges when $x\to x_i$ and $h'''(x_{r+})>0$ because $h'(x_{r+})$ is a minimum, so point $x_2$ is also discarded. On the contrary, $x_3$ is in itself a valid point that gives viability to the model, since by Eq. (\ref{limithprimainfty}) and (\ref{conditionlimit}),
\begin{equation}\label{limitmx3}
    \lim_{x\to\infty}m=+0,
\end{equation}
and 
\begin{equation}\label{derivativeinfty}
    m'(r)\overset{x\to \infty }{=}0.
\end{equation}
Since $0<1+h'(x)$, for $0<x<x_i$, $h''(x)<0$, and $m(x)<0$, and simultaneously for $x>x_i$, $m(x)>0$, therefore Eq. (\ref{limitmx3}) expresses that $m(x)$ should be flattened towards zero at infinity. On the other hand, $r(x)>-1$ for $0<x<x_2$, then it will have a minimum and will tend to $-1$ at infinity, Fig. (\ref{fig2}).
The behavior of $m(r)$ in the phase space can now be drawn as shown in Fig. \ref{fig3}, where $m'(r)$ is also observed. It should be noted that $r'(x)=r(1+m+r)/x$, so that the maximal points of $r$ are found when $m=-1-r$, that is, the points where $m'(r)|_{x}$ diverges. Since $r=-1$ in $x_1$, $x_2$ and $x_3$, it has two maximum points, $x_A$ and $x_B$ in Fig. \ref{fig2}, and by definition, in these points the derivative is infinite, shown in Fig. \ref{fig3}, as $(r_A, m_A)$ and $(r_B,m_B)$.

In the next section a differential equation for $h(x)$ will be constructed from its geometry, considering the parity of the functions $h'(x)$ and $h''(x)$ and multiplying them by functions so that their roots coincide.

\section{Starobinsky type models}\label{sec:5}
Since $h(x)$ is an even function, $h'(x)$ and $h''(x)$ are odd and even functions respectively, as can be seen in Fig. \ref{fig1}, so it can be inferred that the function $x h''(x)$ will be odd with roots in $x=0$ and $x=x_i$. Simultaneously, when multiplying $h'(x)$ by the factor $(x_i^2-x^2)$, the same intervals of increase and decrease of $x h''(x)$ are obtained, besides the same roots, for both negative and positive $x$. Let us assume that $h'(x)$ and $h''(x)$ can be related by
\begin{equation}\label{suposi}
    q(1+p(x))x h''(x)=\left(x_i^2-x^2\right)h'(x),
\end{equation}
where $q$ is some constant and $p(x)$ is a function that is linearly independent of $h(x)$ and also even so that the left member of Eq. (\ref{suposi}) is odd, moreover
\begin{equation}\label{limitp}
    \lim_{x\to0}p(x)=0,
\end{equation}
and
\begin{equation}\label{limitpinfty}
    \lim_{x\to\infty}\frac{1}{1+p(x)}=\lim_{x\to\infty}\frac{q}{x^2-x_i^2}=0.
\end{equation}
since $q$ is a constant, it is possible to evaluate it at some limit, for example
\begin{equation}
    q=\lim_{x\to0}\frac{(x_i^2-x^2)h'(x)}{\left(1+p(x)\right)xh''(x)}=\lim_{x\to0}\frac{x_i^2}{1+p(x)}=x_i^2,
\end{equation}
thus Eq. (\ref{suposi}) can be rewritten as
\begin{equation}
    \left(1+p(x)\right)\frac{1}{x}h''(x)+\left(\frac{1}{x_i^2}-\frac{1}{x^2}\right)h'(x)=0,
\end{equation}
and integrating it
\begin{equation}
    \frac{h(x)}{x_i^2}+\frac{h'(x)}{x}+\int \frac{1}{x}p(x)h''(x)dx=0,
\end{equation}
where it can be seen why $1+p(x)$ was chosen in Eq. (\ref{suposi}) rather than $p(x)$. The last integral can be evaluated by parts twice, so that it does not depend on $h''(x)$, but on $h(x)$ and $h'(x)$, or equivalently for the purpose of reducing the order of the differential Eq. (\ref{suposi}).
\begin{widetext}
\begin{equation}
    \int \frac{1}{x}p(x)h''(x)dx=\frac{1}{x}p(x) h'(x)+h(x) \left(\frac{p(x)}{x^2}-\frac{p'(x)}{x}\right)+\int \frac{h(x)}{x^3} \left(x^2 p''(x)-2 x p'(x)+2 p(x)\right) dx,
\end{equation}
\end{widetext}
then $p(x)$ can be chosen as the solution of the differential equation
\begin{equation}
    x^2 p''(x)-2 x p'(x)+2 p(x)=0,
\end{equation}
that is
\begin{equation}
    p(x)=x(\alpha+\beta x),
\end{equation}
with $\alpha$ and $\beta$ constants, but for $p(x)$ to be an even function, $\alpha=0$, so the equation is reduced to
\begin{equation}\label{dif1}
    \left(\frac{1}{x}+\beta x\right)h'(x)+\left(\frac{1}{x_i^2}-\beta\right)h(x)+c_1=0,
\end{equation}
whose solution is
\begin{equation}\label{model1}
    h(x)=\frac{k}{R_0}\left[1-\left(1+\beta x^2\right)^{\frac{1}{2}-\frac{1}{2\beta x_i^2}}\right]-\lambda,
\end{equation}
where the integration constants were found by Eq. (\ref{limith0}) and (\ref{limithinfty}) for $\beta>0$ and $\beta x_i^2<1$, which, in turn, requires that the power must be negative,
\begin{equation}
    \frac{1}{2}\left(1-\frac{1}{\beta x_i^2}\right)=-m,
\end{equation}
with $m>0$, or analogously, when $x_i=(2m-1)^{-1/2}$ and $\beta=1$,
\begin{equation}\label{staro}
    h(x)=\frac{k}{R_0}\left[1-\left(1+x^2\right)^{-m}\right]-\lambda,
\end{equation}
so without loss of generality, it can be concluded that Eq. (\ref{model1}) actually represents Starobinsky's model \cite{Starobinsky_2007}. In this way it is easy to find the value of the constant $c$, Eq. (\ref{limithdobleprima0}),
\begin{equation}
    c=\frac{2km}{R_0},
\end{equation}
and therefore
\begin{equation}\label{starogeneral}
    y(x)=x+\frac{k}{R_0}\left[1-\left(1+x^2\right)^{-m}\right],
\end{equation}
where $m, k, R_0$ are parameters. In the next section, a generalized Hu-Sawicki model will be established through a similar procedure.
\section{Hu-Sawicki type models}\label{sec:6}
Equation (\ref{suposi}) relates $h'(x)$ and $h''(x)$ in a geometric way through the inflection point $x_i$, that is by the root of $h''(x)$, provided by the difference $x_i^2-x^2$ in the right member of the equation. However, in the more general case it is possible to write
\begin{equation}\label{suposi2}
    ts(x)xh''(x)=(x_i^r-x^r)h'(x),
\end{equation}
where $t$ is constant, $r>0$ is an even number and $s(x)$ is a continuous even function that, as in the previous section, will be requested linearly independent of $h(x)$, satisfying 
\begin{equation}
    \lim_{x\to x_i}s(x)=-\frac{r x_i^{r-2}h'(x_i)}{th'''(x_i)},
\end{equation}
and
\begin{equation}
    \lim_{x\to\infty}\frac{1}{s(x)}=0.
\end{equation}
Note that Eq. (\ref{suposi2}) can be integrated as

\begin{widetext}
\begin{equation}
    \frac{1}{1+r}x\left[x^r-(1+r)x_i^r\right]h'(x)+\int\left[ts(x)-\frac{x^r}{1+r}+x_i^r\right]xh''(x)dx+c_1=0,
\end{equation}
and integrating by parts twice the last integral
\begin{eqnarray}
    \frac{1}{1+r}x\left[x^r-(1+r)x_i^r\right]h'(x)+\left(ts(x)-\frac{x^r}{1+r}+x_i^r\right)xh'(x)+\left[x^r-x_i^r-t\left(s(x)+xs'(x)\right)\right]h(x)-&&\nonumber\\
    \int\left[rx^{r-1}-t\left(2s'(x)+xs''(x)\right)\right]h(x)dx+c_1&=&0,
\end{eqnarray}
\end{widetext}
for a null integrand,
\begin{equation}
    s(x)=\frac{x^r}{t(1+r)}-\frac{\alpha}{x}+\beta,
\end{equation}
with $\alpha=0$ to allow the function to be even, thus Eq. (\ref{suposi2}) can be written as
\begin{equation}\label{ineq2}
    \left(\frac{x^r}{1+r}+t\beta\right)xh'(x)-(x_i^r+t\beta)h(x)+c_1=0,
\end{equation}
whose solution is
\begin{equation}
    h(x)=\frac{k}{R_0}x^{1+\frac{x^r}{t}}\left[x^r+(1+r)t\right]^{-\frac{x_i^r+t}{r t}}-\lambda,
\end{equation}
where $\beta$ was absorbed in $t$, the constants were established from conditions (\ref{limith0}) and (\ref{limithinfty}), and by (\ref{condition1}), $k<0$. However, for this function to be even, the ratio $x_i^r/t$ must be an odd number, and since the parameter $r$ is even, they can be related by means of 
\begin{equation}
    x_i^r=(n r-1)t,
\end{equation}
where $n$ is a natural number, and consequently, the function can be expressed as
\begin{equation}\label{model2}
    h(x)=\frac{k}{R_0}\left[1+(1+r)t x^{-r}\right]^{-n}-\lambda,
\end{equation}
with which for $nr>2$ and $t>0$, $c=0$. In this scenario Eq. (\ref{xdefinition}) is now expressed as
\begin{equation}\label{husawgeneral}
    y(x)=x+\frac{k}{R_0}\left[1+(1+r)t x^{-r}\right]^{-n}.
\end{equation}
As a particular case, when $n=1$ and defining
\begin{equation}
    c_1=-\frac{c_2k}{R_0},\qquad\text{and}\qquad c_2=\frac{1}{(1+r)t},
\end{equation}
the inflection point is obtained at
\begin{equation}
    x_i^r=\frac{r-1}{c_2(r+1)},
\end{equation}
and
\begin{equation}
    h(x)=-\lambda-\frac{c_1x^r}{1+c_2x^r},
\end{equation}
so
\begin{equation}
    y(x)=x-\frac{c_1x^r}{1+c_2x^r},
\end{equation}
which is the Hu-Sawicki model \cite{Hu_2007}, where $c=0$. Nevertheless, because $r>0$, it is not possible to obtain Starobinsky's model from Eq. (\ref{model2}) and therefore Eq. (\ref{model1}) and (\ref{model2}) represent different models, however, these models are part of a more general class of models, as will be seen in the next section.
\section{Hypergeometric models}\label{sec:7}
The similarities of models given by equations (\ref{model1}) and (\ref{model2}) can be found in the form of their differential equations, as well as in the possible values of their respective parameters. To see this, Eq. (\ref{dif1}) is rewritten as
\begin{equation}\label{eq61}
    \frac{1+\beta x^2}{(1+2m-\beta)x}h'(x)+h(x)+\lambda-\frac{k}{R_0}=0,
\end{equation}
and Eq. (\ref{suposi})
\begin{equation}\label{eq62}
    \left(1+\beta x^2\right)xh''(x)+\left[(1+2m)x^2-1\right]h'(x)=0,
\end{equation}
multiplying Eq. (\ref{eq61}) for $n r$ and making $\beta=1$, $v=-1$ and $r=-2$, these equations can be combined as
\begin{multline}\label{combi1}
    (v-x^r)x^2h''(x)+\left[v\left(1-(m+n)r\right)+(1+nr)x^r\right]xh'(x)\\
    +mnr^2vh(x)+2mnr\left(\lambda-\frac{k}{R_0}\right)=0.
\end{multline}
In the the same manner, it is possible to express Eq. (\ref{ineq2}) and (\ref{suposi2}), respectively, as
\begin{equation}
    \left(\frac{x^r}{t(1+r)}+1\right)xh'(x)-n rh(x)-n r\lambda=0,
\end{equation}
and
\begin{equation}
    \left(\frac{x^r}{t(1+r)}+1\right)xh''(x)+\left(\frac{x^r}{t}+1-n r\right)h'(x)=0,
\end{equation}
\begin{widetext}
or combined as
\begin{equation}\label{combi2}
(v-x^r)x^2h''(x)+\left[v\left(1-(m+n)r\right)+((m-1)r-1)x^r\right]xh'(x)+mnr^2vh(x)=0,
\end{equation}
where $v=-t\beta(1+r)$ and $\lambda=0$. Therefore, a generalization of Eq. (\ref{combi1}) and (\ref{combi2}) can be made as follows,
\begin{equation}\label{combi3}
    (v-x^r)x^2h''(x)+\left[v\left(1-(m+n)r\right)+((u-1)r-1)x^r\right]xh'(x)+mnr^2v\left(h(x)+\lambda-\frac{c}{2m}\right)=0,
\end{equation}
\end{widetext}
where $u$ is a parameter that can be adjusted according to the type of model, when $u=n$ and $r=-2$, the Starobinsky type model is obtained, Eq. (\ref{staro}), and when $u=m$, the Hu-Sawicki one, Eq. (\ref{model2}), is found. Now with the variable change
\begin{equation}
    z=v x^{-r},
\end{equation}
it is realized that Eq. (\ref{combi3}) is in effect the hypergeometric equation
\begin{equation}\label{hipergeq}
    (1-z)zg''(z)+(u-(1+m+n)z)g'(z)-mng(z)=0,
\end{equation}
where
\begin{equation}
    g(z)=h(z)+\lambda-\frac{c}{2m},
\end{equation}
by choosing the constants appropriately, according to equations (\ref{staro}) and (\ref{model2}), the solution can be written as (for $m,n,u,v,r\neq0$)
\begin{equation}\label{hipergeo}
    h(z)=\left(\frac{k}{R_0}-\frac{c}{m}\right)\,_2F_1\left(m,n,u;z\right)+\frac{c}{2m}-\lambda,
\end{equation}
together with the condition of existence of the inflection point, given by the algebraic equation
\begin{equation}\label{inflectionpoint}
    \frac{\, _2F_1\left(m+1,n+1;u+1;z_i\right)}{\, _2F_1\left(m+2,n+2;u+2;z_i\right)}=\frac{(m+1) (n+1) r}{(r+1) (u+1)z_i}.
\end{equation}
If this equation ensures the existence of a single point of inflection, it remains to analyze the domain of the parameters in which $h(x)$ is viable. When $r>0$, the model naturally satisfies limits Eq. (\ref{limithprimainfty}) and Eq. (\ref{limithdobleprinfini}), however to satisfy the limits (\ref{limith0}) and (\ref{limithinfty}), $c=0$, and at the same time, by the series expansion of $h'(x)$ and $h''(x)$, it is observed that $m>2/r$, $n>2/r$, $v\neq0$, $m-n\notin \mathbb{Z}$ and $u\neq w$ with $w\in\mathbb{Z}$ and $w\leq0$, for the model to meet limits (\ref{limithprima0}) and (\ref{limithdobleprima0}). In addition, Euler's integral representation of the hypergeometric function allows to further restrict the parameters of the model according to Eq. (\ref{condition1}), since for $x>0$, $-1<n<u$ and $v<0$ (with $R_0=1$)
\begin{equation}\label{hintegral}
    h'(x)=-\frac{k m r v x^{m r-1}\Gamma (u)}{\Gamma (n) \Gamma (u-n)}\int_0^1\frac{(1-t)^{u-n-1}t^n}{(x^r-v t)^{m+1}}dt,
\end{equation}
where $\Gamma(x)$ is the Gamma function. Due to the positive integrand and the integral definition interval, the integral is positive, and because
\begin{equation}
    \frac{\Gamma (u)}{\Gamma (n) \Gamma (u-n)}>0,
\end{equation}
for $n<u$, then for $h'(x)<0$, $k<0$. Simultaneously, in order that $h'(x)>-1$,
\begin{equation}\label{lastcondition}
    \frac{u(r+1)(u+1)x_i^{2 r+1}}{|k|r^2 v^2mn(m+1)(n+1)}>\, _2F_1\left(m+2,n+2;u+2;\frac{v}{x_i^r}\right),
\end{equation}
where $x_i$ is the inflection point, obtained from Eq. (\ref{inflectionpoint}). 
Numerically it is found that for $0<r<2$, $m\geq1$, $n\geq1$, $n<u<n+1$, $v<-1$,
\begin{equation}
    x^{r+1}>\, _2F_1\left(m+1,n+1;u+1;v x^{-r}\right),
\end{equation}
so that if $u>k m n r v$, then $h'(x)>-1$.

Alternatively, a sufficient condition for $h'(x)>-1$ is
\begin{equation}
    \frac{\left(1+\frac{1}{r}\right)_{w-1} (u)_w}{r \left| k v^w\right| (m)_w (n)_w}>\frac{\, _2F_1\left(m+w,n+w;u+w;\frac{v}{x_i^r}\right)}{x_i^{r w+1}},
\end{equation}
where $(a)_w$ is the Pochhammer symbol.

On the other hand, when $r<0$, for limit Eq. (\ref{limith0}) to be satisfied, $c=2km/R_0$, whereas for limits (\ref{limithinfty}), (\ref{limithprimainfty}) and (\ref{limithdobleprinfini}), $m>0$, $n>0$ and, as in the previous case, $u\neq w$. Likewise, when $r\leq-2$, limit Eq. (\ref{limithprima0}) is satisfied \footnote{When $r\in(0,2)$, $r\neq-1$, both limits Eq. (\ref{limithprima0}) and Eq. (\ref{limithdobleprima0}) are indeterminate, and for $r=-1$, $\lim_{x\to0}h'(x)=-\frac{kmnv}{uR_0}\neq0$}, however when $r=-2$, 
\begin{equation}
    \lim_{x\to0}h''(x)=-\frac{2k m n v}{u R_0},
\end{equation}
so that $u=-n v$, which in turn implies that $v<-1$. When $r<-2$, limit given by Eq. (\ref{limithdobleprima0}) is fulfilled. Note that although the value of the constant $c$ depends on the sign of $r$, the Eq. (\ref{hintegral}) is still valid in this case, $r\leq-2$, so the restrictions on the parameters that were made previously, i.e. $n<u$ and $k<0$, remain valid.

Finally, the hypergeometric model can be expressed using Eq. (\ref{xdefinition}),
\begin{equation}\label{general}
    y(x)=x+\left(\frac{k}{R_0}-\frac{c}{m}\right)\,_2F_1\left(m,n,u;\frac{v}{x^r}\right)+\frac{c}{2m},
\end{equation}
and contains the generalized Starobinsky type (\ref{starogeneral}) and Hu-Sawicki (\ref{husawgeneral}) type models, since when $u=n$, $r=-2$, $v=-1$ and $c=2km/R_0$, the first one is obtained, and when $u=m$, $v=-(1+r)t$ and $c=0$, the second one is obtained, therefore Eq. (\ref{general}) can be considered as a generalization of these models.

\section{Concluding remarks}\label{sec:discu}
A cosmologically viable hypergeometric model in the $f(R)$ gravity theory has been constructed from the assumption of the existence of an inflection point of the $f(R)$ curve, the viability conditions in the $(m, r)$ plane and such that it reproduces $\Lambda$CDM at some limit. This last quality was used to express the limits of the model, written in terms of the dimensionless variable $x$, as $y(x)=x+h(x)+\lambda$, where $h(x)$ represents the deviation of the model from GR. From the geometric point of view, the existence of the inflection point, $x_i$, besides the decreasing monotonicity of $h(x)$, ensure that limits (\ref{limith0}) and (\ref{limithinfty}) are satisfied, allowing to consider the model as a perturbation around $\Lambda$CDM, and at the same time when $x\to\infty$, the conditions $r=-1$, $m=+0$ and $m'(-1)=0$ are met, see figures (\ref{fig2}) and (\ref{fig3}), enabling $y(x)$ to have a matter domination epoch. The physical interpretation of $x_i$ is indeed to allow the model to have an asymptotic behaviour towards $\Lambda$CDM.

The geometrical conditions imposed by $x_i$, both on the function $h(x)$ and its derivatives, was used to construct a differential equation in such a way that the roots of $x h''(x)$, modulated by a function $p(x)$, coincide with a term $h'(x)$ multiplied by the factor $(x^2-x_i^2)$. This differential equation was integrated and the function $p(x)$ was chosen to allow the integrand to be expressed as an exact differential. The solution, Eq. (\ref{starogeneral}) corresponds to Starobinsky's 2007 model \cite{Starobinsky_2007}. Through a similar procedure, but this time expressing the factor as $(x^r-x_i^r)$, being $r$ a parameter of the model, a differential equation was constructed whose solution, Eq. (\ref{husawgeneral}), corresponded to a generalization of Hu-Sawicki model \cite{Hu_2007}.
\\
It was found that the differential equations of each model in effect belonged to a particular case of the hypergeometric differential equation, and as a result the hypergeometric model, Eq. (\ref{general}), could be established. This model depends on five parameters $m, n, r, u$, and $v$, however, the equation for the inflection point, (\ref{inflectionpoint}), represents a constriction of the model, since it is a necessary condition for its viability, reducing the number of parameters to four. Moreover the constant $k<0$ and the value of $c$, in a concrete way, can be generally determined from the classification in the two sub-models, Hu-Sawicki type ($r>0$): $c=0$; and Starobinsky type ($r<0$): $c=2km/R_0$.
\\
When $r>0$ and $x>0$, for the model to satisfy limits (\ref{limith0}) to (\ref{limithdobleprinfini}), as well as condition Eq. (\ref{condition1}), the parameters must fulfill $m>2/r$, $u>n>2/r$, $v<0$, and $m-n\notin \mathbb{Z}$, in addition to Eq. (\ref{lastcondition}). For example, when $0<r<2$, $m\geq1$, $n\geq1$, $n<u<n+1$, $v<-1$, $u>k m n r v$, the hypergeometric model is cosmologically viable.
\\
At the other hand, only for $r\leq2$, $m>0$ and $u>n>0$, the hypergeometric model is viable. Specifically when $r=-2$, it is found that $u=-nv$ and $v<-1$.

The main quality of the hypergeometric model is that it encompasses a family of functions that have an inflection point and at the same time mimics the $\Lambda$CDM model, examples of which are the well-known Starobinsky and Hu-Sawicki models. The hypergeometric model proposed here depends on four free parameters, offering the possibility of having greater freedom of adjustment according to the restrictions offered by observational data at both cosmological and local scales. To carry out this objective, it should be noted, the appropriate computational tools are needed as indicated in Ref. \cite{hough2020viability}, however, the outlook for achieving this goal is encouraging, since in the near future modified gravity theory could be tested by major advances in observational techniques in high curvature scenarios such as black holes or neutron stars, where $f(R)$ can play an important role in the dynamics of spacetime, and its effects could be appreciated.

\appendix

\begin{widetext}
\section{Limit of $m'(r)$}
Note that the derivative of $m(r)$ can be expressed as
\begin{equation}
    m'(r)|_{x}=-\frac{\left(\lambda+x+h(x)\right)^2}{\lambda+h(x)} \frac{\left(1+h'(x)-x h''(x)\right)h''(x)+\left(1+h'(x)\right)x h'''(x)}{\left(1+h'(x)\right)^2 \left[\left(1+h'(x)\right) \left(1-\frac{x h'(x)}{\lambda+h(x)}\right)+x h''(x)+\frac{x^2 h''(x)}{\lambda+h(x)}\right]},
\end{equation}
such that, by Eq. (\ref{limith0}), (\ref{limithprima0}) and (\ref{limithdobleprima0}),
\begin{equation}
    \lim_{x\to 0}\frac{\left(\lambda+x+h(x)\right)^2}{\lambda+h(x)}=\lim_{x\to 0}\frac{2\left(1+h'(x)\right)^2+2\left(\lambda+x+h(x)\right)h''(x)}{h''(x)}=\frac{2}{c},
\end{equation}
\end{widetext}
\begin{equation}
    \lim_{x\to0}\frac{x h'(x)}{\lambda+h(x)}=\lim_{x\to0}\frac{2h''(x)+xh'''(x)}{h''(x)}=2,
\end{equation}
and
\begin{equation}
    \lim_{x\to0}\frac{x^2 h''(x)}{\lambda+h(x)}=\lim_{x\to0}\frac{2h''(x)+4xh'''(x)+x^2h^{(4)}(x)}{h''(x)}=2,
\end{equation}
therefore
\begin{equation}
    \lim_{x\to0}m'(r)|_x=-2.
\end{equation}

\nocite{*}

\bibliography{apssamp}

\end{document}